\UseRawInputEncoding
\documentclass[aps,prb,groupedaddress,twocolumn,showpacs]{revtex4-1}
\usepackage[dvips]{graphicx}
\usepackage{dcolumn}
\usepackage{bm}
\usepackage{amssymb}
\usepackage{epstopdf}
\usepackage{amsmath}
\usepackage{color}
\usepackage[dvipdfmx]{hyperref}
\hypersetup{
 setpagesize = false,
 colorlinks = true,
 citecolor = blue,
 linkcolor = blue,
 urlcolor = blue,
 }
\begin{document}

\title{Modulating magnetic and electronic properties of lead-apatite Pb$_{10}$(PO$_4$)$_6$O via different transition metal doping}

\author{Bingxin Liu$^{1,2}$}
\author{Tong Liu$^{2}$}
\author{Xun-Wang Yan$^{2}$}\email{yanxunwang@163.com}
\author{Zong-Liang Li$^{1}$}\email{lizongliang@sdnu.edu.cn}
\date{\today}

\affiliation{$^{1}$School of Physics and Electronics, Shandong Normal University, Jinan 250358, China}
\affiliation{$^{2}$College of Physics and Engineering, Qufu Normal University, Qufu, Shandong 273165, China}
\begin{abstract}

Based on the first-principles electronic structure calculations, we investigate the magnetic and electronic properties of transition-metal-doped Pb$_{10}$(PO$_4$)$_6$O with the Pb atom substituted by various transition metals. The $3d$ orbitals of the doped transition metals are distributed near the Fermi energy and exhibit strong spin-polarization, resulting in the local magnetic moment. With the doped metal varying from V, Cr, Mn, Fe, Co, Ni, Cu, to Zn, the moment changes gradually from 3.0, 4.0, 5.0, 4.0, 3.0, 2.0, 1.0, to 0 $\mu_B$, while the electronic structure is modulated progressively due to the continuous change of total charge. Moreover, the flat bands near the Fermi energy are also found in the V, Cr, and Fe-doped lead-apatite, similar to the Cu-doped system. Our work not only provides the critical comparative information for investigating the Cu-doped lead-apatite, but also suggest a category of diluted magnetic semiconductor.

\end{abstract}


\maketitle

\section{Introduction}
The realization of superconductivity at room temperature and ambient pressure has always been a long-standing target in physics, chemistry, and materials science. Although some great achievements have been made in the superconductivity field in the past years \cite{Drozdov2015,Somayazulu2019,Troyan2021}, there are still many huge challenges both in the theoretical and experimental researches.
Very recently, the superconductivity in Pb$_{10-x}$Cu$_x$(PO$_4$)$_6$O (0.9 $\leq$ x $\leq$ 1.1)\cite{Lee2023,Lee2023a} with the critical temperature at 400 K was reported by Lee $et ~al.$. The results is so shocking that many researchers wonder the experimental results is really related to the high temperature superconductivity.
Many experimental and theoretical efforts have been made to synthesise the LK-99 sample and replicate the claimed superconductivity, and one result after another comes up shortly.
Kumar $et ~al.$ reported their polycrystalline samples of Pb$_{10-x}$Cu$_x$(PO$_4$)$_6$O \cite{Kumar2023,Kumar2023a}, which displayed the similar X-ray diffraction (XRD) spectra to the ones from Lee's results \cite{Lee2023}, but the occurrence of superconductivity was not approved by the magnetism measurements.
Liu $et ~al.$ fabricated the Pb$_{10-x}$Cu$_x$(PO$_4$)$_6$O samples \cite{liu2023}, and the samples exhibit semiconductor-like transport behavior with a large room-temperature resistivity of 1.94 $\times$ 10$^{4}$ $\Omega \cdot$cm and no magnetic levitation was observed.
Then, another research team from Southeast University in China stated that they found zero resistance in Pb$_{10-x}$Cu$_x$(PO$_4$)$_6$O at 110 K, far below room temperature\cite{Hou2023}.
The magnetic levitation with larger angle than Sukbae Lee's sample was realized in the article from Wu $et ~al.$ \cite{Wu2023}.
Guo $et ~al.$ observed the half levitation of the samples and explained that the levitation is related to the weak and soft ferromagnetism, not the Meissner effect \cite{Guo2023}.
An important report from Zhu $et ~al.$ argued that the so-called superconducting behavior in LK-99 is most likely due to a reduction in resistivity caused by the first order structural phase transition of Cu$_2$S at around 385 K, from the $\beta$ phase to the $\gamma$ phase \cite{Zhu2023}.
In the updated article of Wu $et ~al.$ \cite{Wu2023a}, a steep resistance jump at 387 K was found and they thought that it is difficult to identify whether the jump come from the superconducting component or impurity in LK-99 samples.

In the theoretical researches, Griffin first posted the theoretical analysis of the electronic structure of Pb$_{10-x}$Cu$_x$(PO$_4$)$_6$O and the flat bands near the Fermi energy, which is usually related to the strong electronic correlation and lead to the novel physics \cite{Griffin2023}.
Several other theoretical studies also suggested the presence of flat bands \cite{lai2023,Si2023,Tao2023}, which is based on the same assumption of crystal structure that Cu substitute the Pb atom at $4f$ Wyckoff site. Apart from these, two first-principles works provide different conclusion that the Cu-doped lead apatite is semiconducting \cite{Sun2023b,Zhang2023}, which is derived from the crystal structure with Cu substituting the Pb atom in $6h$ Wyckoff site.

Up to now, apart from Cu, other $3d$ transition metals (TM) have not been used to prepare the sample in experiments. At the same, there is few theoretical studies to involve other $3d$ metals doping. We know that the doping of Cu and other $3d$ metals are very likely to induce the similar physical phenomenon in the lead apatite. Different $3d$ elements doping can effectively modulate and manipulate the magnetism in modified lead apatite because of the continuous adjustment of $d$ electron number from one to another. At present, the sharp resistivity jump, ferromagnetism, soft ferromagnetism, and diamagnetism have been observed in the different experiments, and the reason of resistivity jump and the true magnetism are still unknown and mysterious. We believe that different $3d$ transition metals doping can provide an important reference and comparison for investigating the Cu-doped lead apatite.
 Therefore, what changes in the magnetic and electronic properties will be brought about by the different $3d$ metals doping is a topic worth paying attention to.

\section{Computational details}
The plane wave pseudopotential method enclosed in Vienna Ab initio simulation package (VASP) \cite{PhysRevB.47.558, PhysRevB.54.11169} is used in our calculations. The generalized gradient approximation (GGA) with Perdew-Burke-Ernzerhof (PBE) formula \cite{PhysRevLett.77.3865} as well as the projector augmented-wave method (PAW) \cite{PhysRevB.50.17953} were adopted in ionic potentials.
The plane wave basis cutoff was 600 eV and the thresholds were 10$^{-5}$ eV and 0.01 eV/\AA ~ for total energy and force convergence.
 For structural optimizations, a mesh of $4\times 4\times 4$ k-points were sampled for the Brillouin zone integration, and the $8\times 8\times 8$ k-mesh for the self-consistent electronic structure calculations.
The GGA + U method with the empirical value of Hubbard U is used to correct the electron correlation effect in the electronic structure calculations \cite{Cococcioni2005}.
\section{Results and discussion}
\subsection{Atomic structure}
\begin{figure}
\begin{center}
\includegraphics[width=8.5cm]{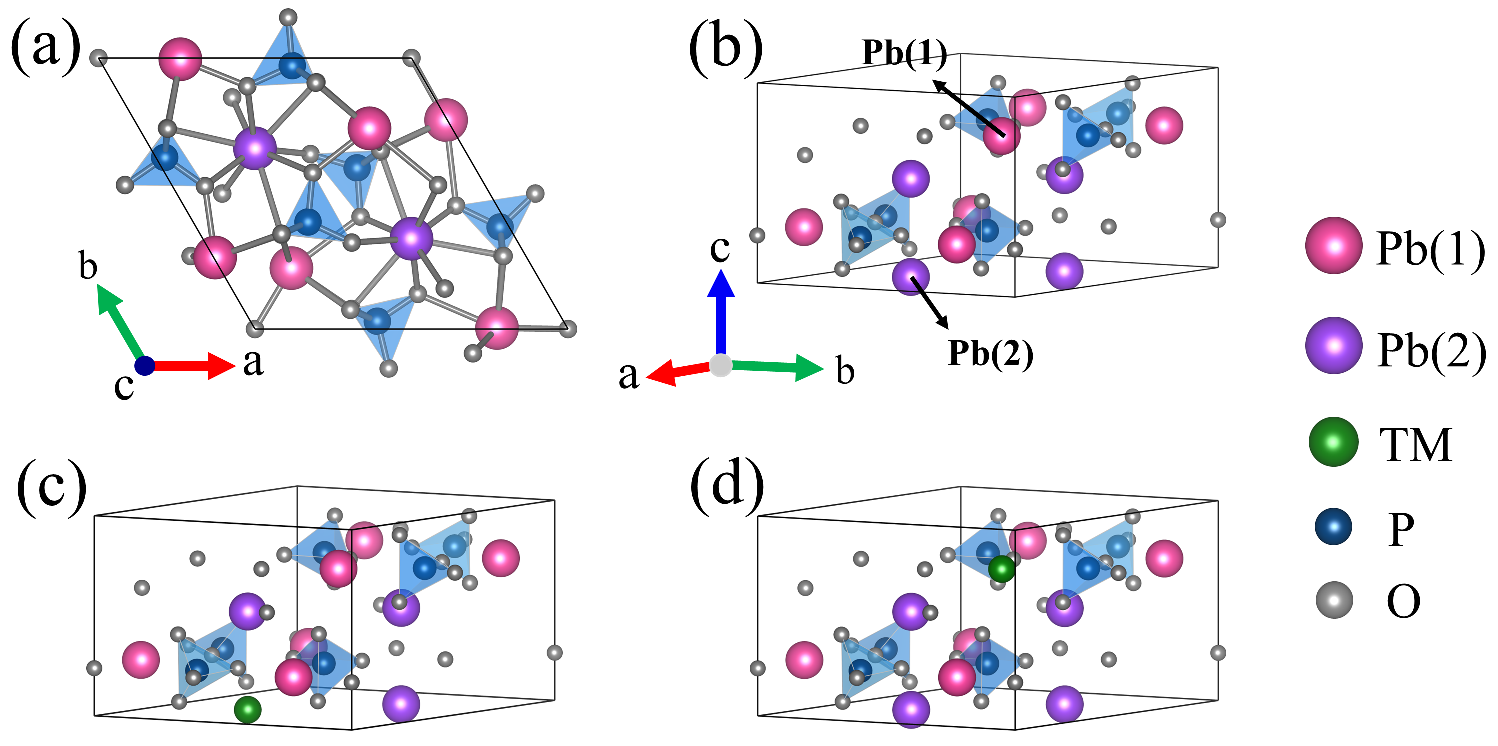}
\caption{Atomic structure of Pb$_{10}$(PO$_4$)$_6$O. (a) Top view, (b) The positions of Pb(1) and Pb(2) from side view, (c) The doped Cu atom to replace one Pb(1) atom, (d) The doped Cu atom to replace one Pb(2) atom.
 } \label{structmodel}
\end{center}
\end{figure}

 The atomic structure of Pb$_{10}$(PO$_4$)$_6$O compound is sketched in Fig. \ref{structmodel}, which crystalizes in the space group $P6_3/m$ (No. 176) with the crystal constants $a$ = 9.8650 \AA~ and $c$ = 7.4306 \AA~\cite{Krivovichev2003}.
In one unit cell of this compound, ten Pb atoms can be divided into two categories, six Pb(1) atoms and four Pb(2) atoms shown in Fig. \ref{structmodel}(b), which are located in the inequivalent Wyckoff sites, $6h$ and $4f$, respectively. The P atom is coordinated by four O atoms and form the PO$_4$ tetrahedron.
The Pb atoms sit in the interstitial space among these PO$_4$ tetrahedrons, where Pb(1) is bonded to five O atoms with the lengths of Pb-O bonds ranging from 2.44 - 2.63 \AA and Pb(2) is bonded to nine O atoms with Pb-O bond distances varying from 2.56 - 2.94 \AA.
The relatively long Pb-O bonds lead to the easy substitution by other metal atoms.
The structural models with transition metal atom replacing the Pb(1) and Pb(2) are displayed in Fig. \ref{structmodel}(c) and (d).
The four $4e$ sites is filled by O atom with $\frac{1}{4}$ occupation. One O atom is placed in the position (0, 0, 0.2658) and keep other three $4e$ positions empty to represent the $\frac{1}{4}$ O atom occupation, and the specific O atom can be found at the $c$ axis of the unit cell in Fig. \ref{structmodel}(b), (c), and (d).

\subsection{Selective substitution of Pb(1) or Pb(2) atom}
In the calculations, the Pb$_{10-x}$Cu$_x$(PO$_4$)$_6$O with x = 1 is adopted to simulate the Pb$_{10-x}$Cu$_x$(PO$_4$)$_6$O (0.9 $\leq$ x $\leq$ 1.1) synthesized in experiments.
The $3d$ TM elements from V, Cr, Mn, Fe, Co, Ni, Cu, to Zn are used to act as the dopants because their $3d$ electrons increase one by one.
Similarly, in the Pb$_{10-x}$TM$_x$(PO$_4$)$_6$O the x value is set to 1.
For the two kinds of Pb atoms in Pb$_{10}$(PO$_4$)$_6$O, Pb(1) at $6h$ site and Pb(2) at $4f$ site, which kind of Pb atom is substituted by the doped $3d$ metal atom is an open question although the Pb(2) at $4f$ site is suggested in the first reports \cite{Lee2023}.
To inspect the detailed structure of doped lead apatite, we carry out the energy calculations with the PBE and GGA + U methods and present the energy values in Table. \ref{energy},
in which the energies for the different substitution with the same TM atom are compared and the lower energy is marked by bold font.
The result is indeed unexpected.
First, the substituted Cu atom is preferred to sit in the Pb(1) at $6h$ site, instead of Pb(2) at $4f$ site proposed by Lee $et~ al.$ and Griffin \cite{Lee2023, Lee2023a, Griffin2023}.
Secondly, the $6h$ site is not always the energetically favorable site for all TM atoms. Namely, V, Mn, Co, Ni, Zn dopants in $4f$ site correspond to the lower energy and more stable configurations, while Cr, Cu, and Fe dopants favorite the $6h$ site.
To show this clearly, we build the Table. \ref{energy-2} and use the checkmarks to mark the low energy structure.

To our knowledge, the selective substitution dependent on the TM element type in the doped compound is rarely reported.

\begin{table*}
	\caption{The energy comparison of Pb$_{9}$TM(PO$_4$)$_6$O compounds with the TM atom in the $6h$ and $4f$ sites . The unit is eV/(formula cell). The energy values are from the PBE and GGA + U calculations.
	}
	\label{energy}
	\renewcommand\tabcolsep{6.5pt} 
\begin{tabular*}{17.80cm}{ccccccccc}
		\hline
		site & V& Cr & Mn & Fe & Co & Ni & Cu & Zn  \\
		\hline
		\textbf{PBE}   &  &  &  &  &  &    \\
        $6h$ &-273.4943 &\textbf{-273.9297} &-273.7853&\textbf{-271.7120}&-269.9535&-268.1519&\textbf{-266.4356}&-265.5985 \\
		$4f$ &\textbf{-274.1305} &-273.7089 &\textbf{-274.0233} & -271.6797 &\textbf{-269.9634}  & \textbf{-268.2527} &-265.9868  & \textbf{-265.8171}  \\
       \textbf{GGA+U} &  &  &  &  &  &   \\
		$6h$ &-272.1214 &\textbf{-273.6348}  &-272.8370  & \textbf{-270.5920} &-268.8695  & -266.7317 & \textbf{-265.4134} & -265.6723  \\
		$4f$ &\textbf{-273.0632} &-272.1465  &\textbf{-273.2158}  & -270.5161 &\textbf{-269.0593}  & \textbf{-267.0040} & -264.7060 & \textbf{-265.7206} \\
		\hline
\end{tabular*}

\end{table*}

\subsection{Electronic structure of Pb$_{9}$TM(PO$_4$)$_6$O}

Although which of $6h$ and $4f$ sites has been identified to be favorable energetically for a specific TM atom dopant, the $6h$ or $4f$ site with higher energy is likely to be filled under some preparation conditions and processes in experiments. To provide more information for other researchers, we present the simulated electronic structures for the two substitution cases.

\begin{figure}
\begin{center}
\includegraphics[width=8.5cm]{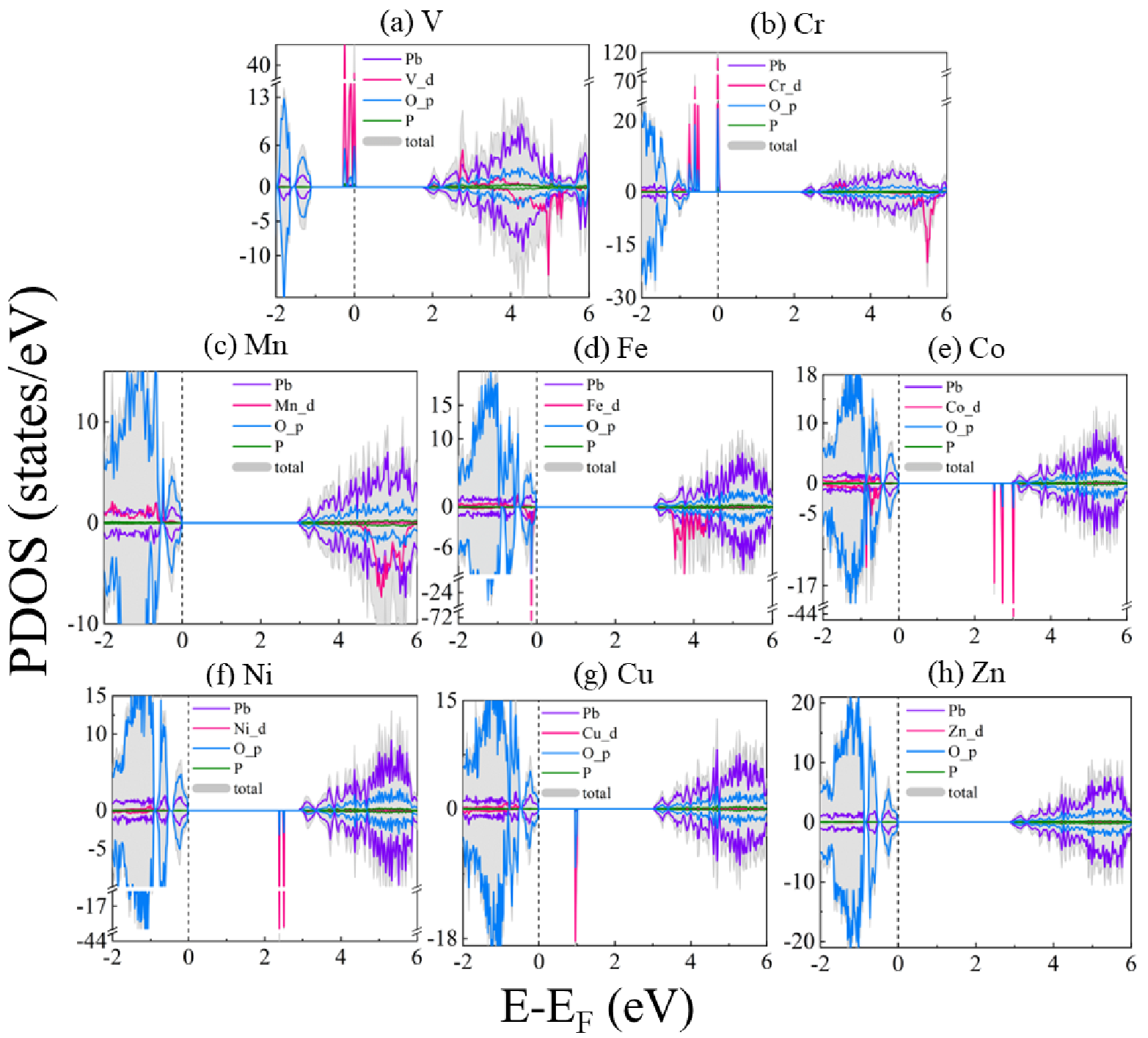}
\caption{Total DOS and Projected DOS of Pb$_{9}$TM(PO$_4$)$_6$O with the different TM atom doping at the $6h$ site.
 } \label{fig2}
\end{center}
\end{figure}

\begin{figure}
\begin{center}
\includegraphics[width=8.50cm]{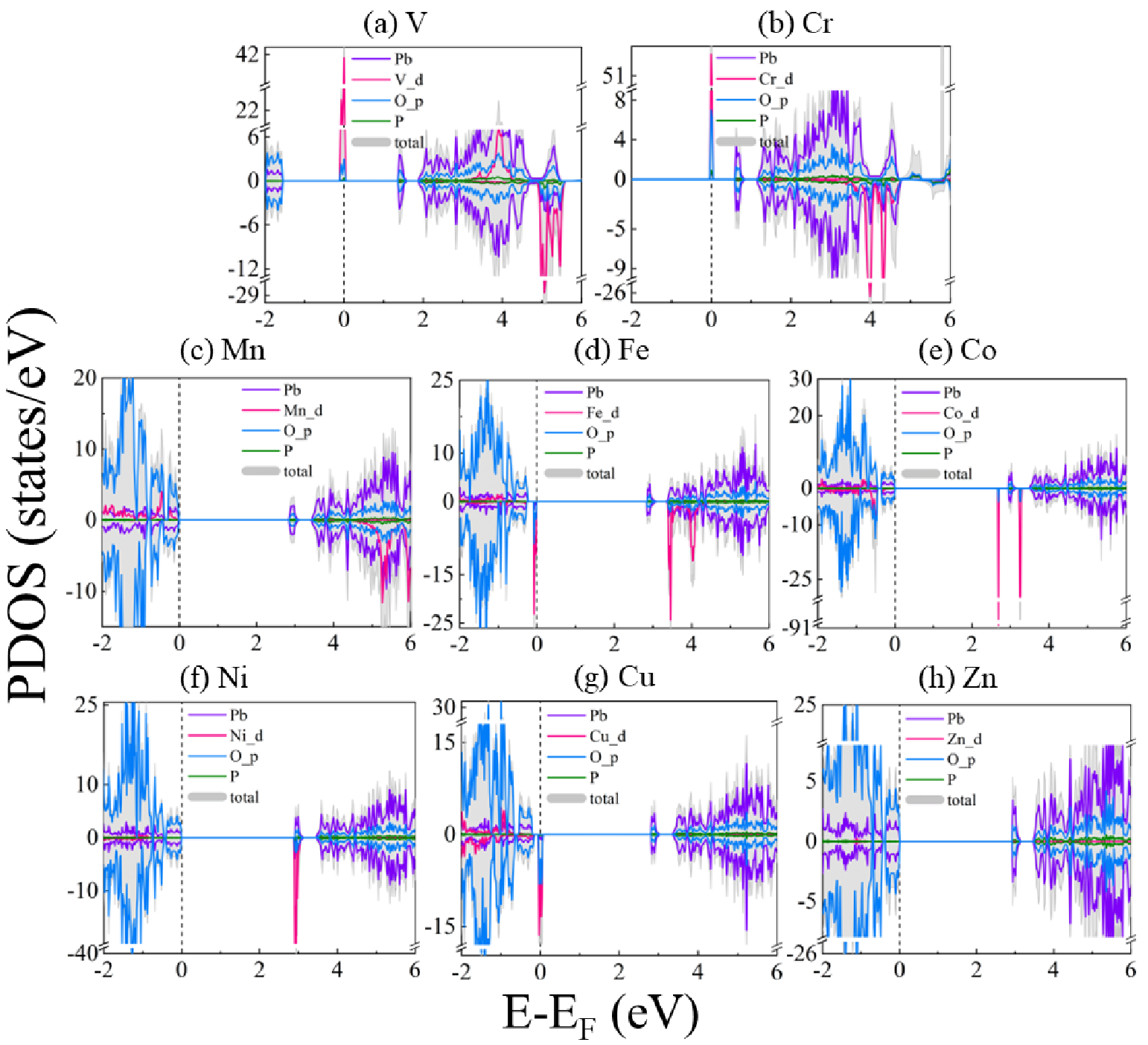}
\caption{Total DOS and Projected DOS of Pb$_{9}$TM(PO$_4$)$_6$O with different TM doping at the $4f$ site.
 } \label{fig3}
\end{center}
\end{figure}

The electronic structure of Pb$_{9}$TM(PO$_4$)$_6$O with the TM replacing Pb(1) and Pb(2) is computed with the GGA + U method with the empirical Hubbard U values. For the Pb(1) replacement case, the total density of states (DOS) and projected DOS on the atomic orbitals are presented in Fig. \ref{fig2}, which shows that they are all insulting with the band gap larger than 2.0 eV for Mn, Fe, Co, Ni, and Zn-doped lead apatite and the gap about 1.0 eV for Cu-doped lead apatite.
For the Pb(2) replacement case, the total and projected DOS are presented in Fig. \ref{fig3}, which indicates that Cu-doped system has a metallic state with a high DOS peak at Fermi energy and others are insulators with the energy gaps ranging from 2.5 to 3.0 eV.
Here, we can see that Cu-doped system is distinct from others whether Cu replacing Pb(1) or Pb(2) atom. This may be the reason why Cu is employed in the reported LK-99 samples.
It is worthy to note the DOS of Fe-doped system in Fig. \ref{fig2}(d) and Fig. \ref{fig3}(d), the high DOS peaks is just below the Fermi energy, in which the doping deviation from x = 1 may shift the sharp peak to Fermi energy and bring about the significant changes in the electronic structure.
As for V and Cr doped systems, there are sharp DOS peaks crossing the Fermi energy, revealing their uniqueness. From the viewpoint of their DOS spectra, V and Cr may be the better dopants to induce the unusual physical phenomenon with respect to Cu element.

\begin{table}
	\caption{For the Pb$_{9}$TM(PO$_4$)$_6$O with TM atom in the $6h$ and $4f$ sites, the structure with lower energy is marked by the signal of checkmark. }
	\label{energy-2}
	\renewcommand\tabcolsep{7.5pt} 
\begin{tabular*}{8.0cm}{ccccccccc}
		\hline
		site& V &Cr & Mn & Fe & Co & Ni & Cu & Zn  \\
		\hline
		$6h$  & & $\checkmark$ &   & $\checkmark$ &  &  & $\checkmark$ &  \\
		$4f$  &$\checkmark$  & &$\checkmark$  &  &$\checkmark$ &$\checkmark$ &  & $\checkmark$ \\
		\hline
\end{tabular*}
\end{table}

\subsection{Flat band feature}
\begin{figure}
\begin{center}
\includegraphics[width=8.50cm]{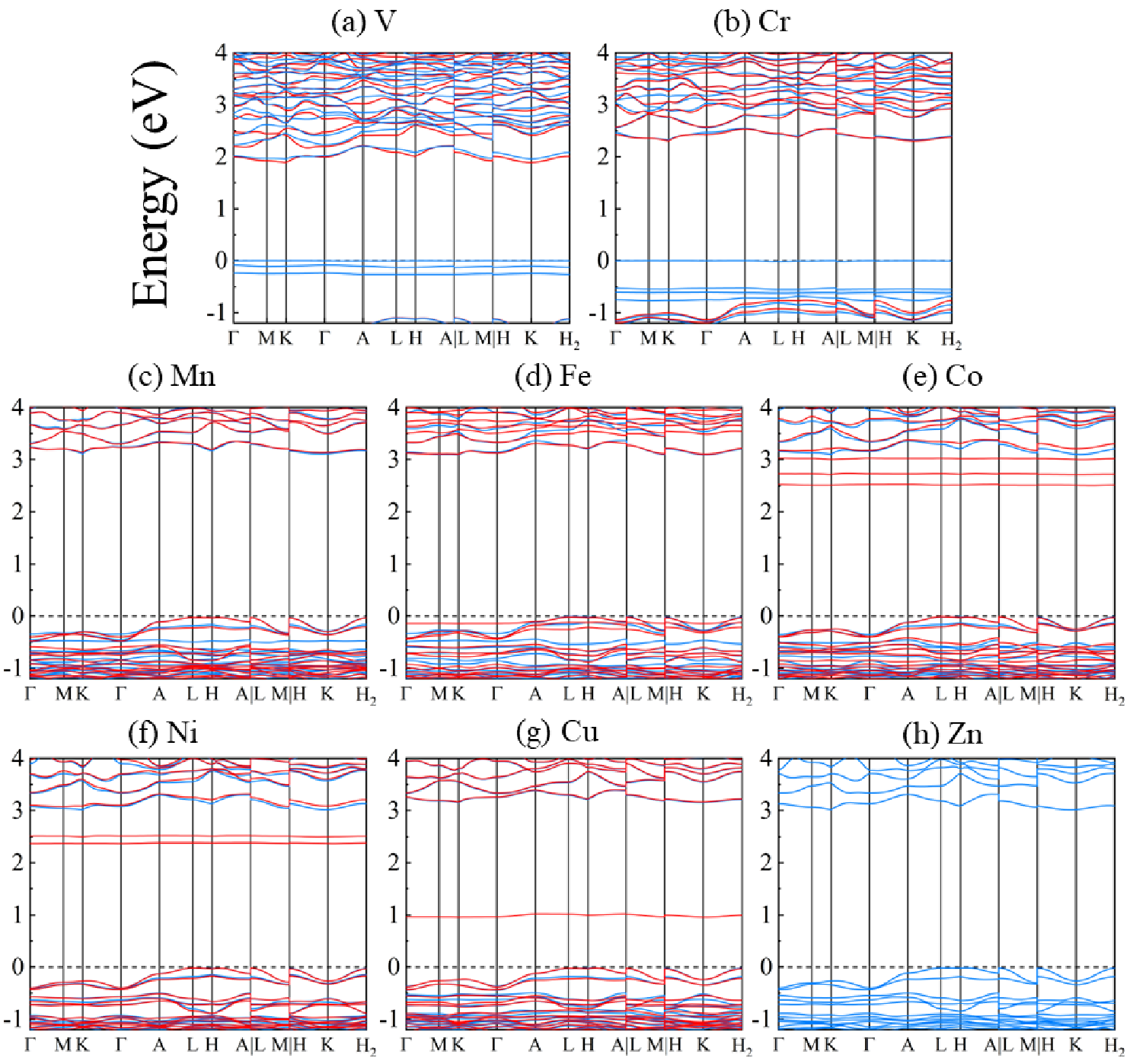}
\caption{Band structures of Pb$_{9}$TM(PO$_4$)$_6$O with the different TM atom doping at the $6h$ site.
 } \label{pb1-u-band}
\end{center}
\end{figure}

\begin{figure}
\begin{center}
\includegraphics[width=8.50cm]{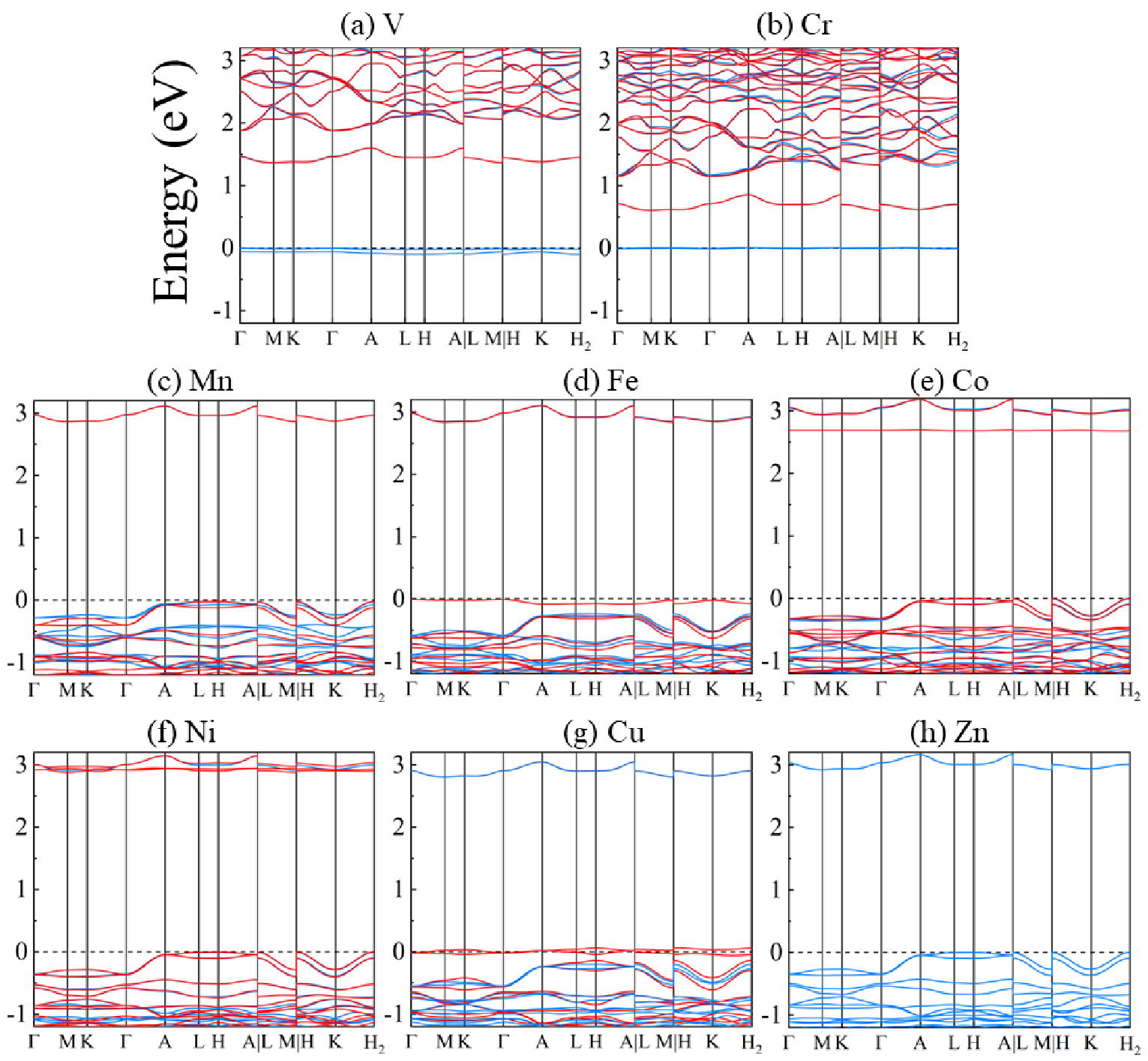}
\caption{Band structures of Pb$_{9}$TM(PO$_4$)$_6$O with the different TM atom doping at the $4f$ site.
 } \label{pb2-u-band}
\end{center}
\end{figure}

Fig. \ref{pb1-u-band} shows the band structures of Pb$_{9}$TM(PO$_4$)$_6$O for the Pb(1) site substituted by V, Cr, Mn, Fe, Co, Ni, Cu, and Zn.
It is noted that there are flat bands crossing the Fermi energy in the band structures of the V- and Cr-doped lead apatite.
The width of the flat bands is less than 0.03 eV, rarely reported in the electronic structure of bulk materials.
For the Pb(2) site substitutions, the band structures of Pb$_{9}$TM(PO$_4$)$_6$O are displayed in Fig. \ref{pb2-u-band}, in which the V-, Cr-, Fe- and Cu-doped systems all exhibit the flat band feature.
Because the magnetic and electrical measurements indicate the Cu-doped Pb$_{10}$(PO$_4$)$_6$O has the complex magnetism and electrical transport properties, we can expect that the electronic structures of V-, Cr-, and Fe-doped lead apatite are also intriguing owning the similar flat band features to Cu-doped system. Therefore, these compounds are worthy of further studies in the experimental and theoretical aspects.

\subsection{Continuous adjustment of magnetic properties}
\begin{figure}
\begin{center}
\includegraphics[width=8.50cm]{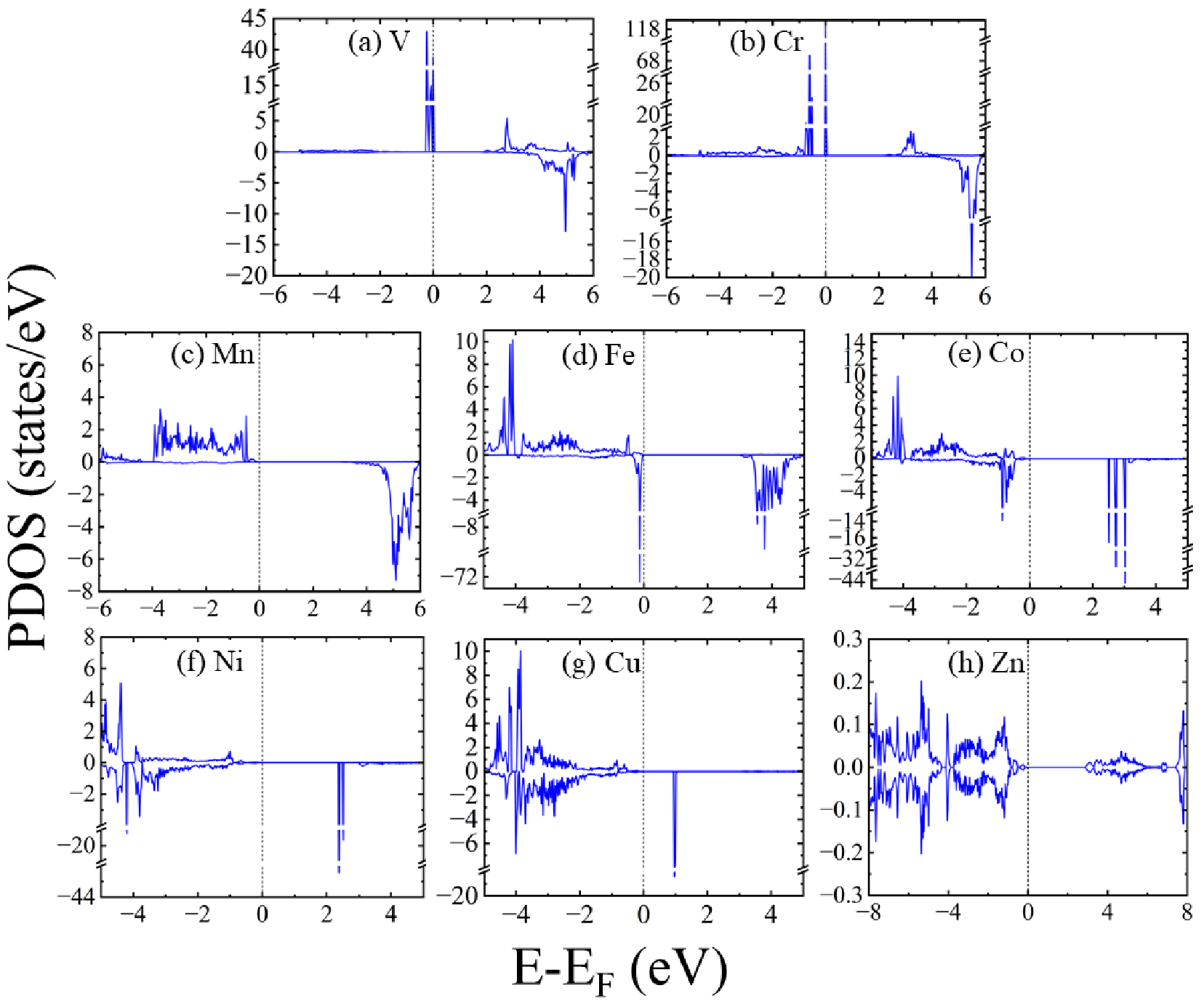}
\caption{Projected DOS on the $3d$ orbitals of TM atom in Pb$_{9}$TM(PO$_4$)$_6$O with the different TM atom in the $6h$ site.
 } \label{fig4}
\end{center}
\end{figure}

\begin{figure}
\begin{center}
\includegraphics[width=8.50cm]{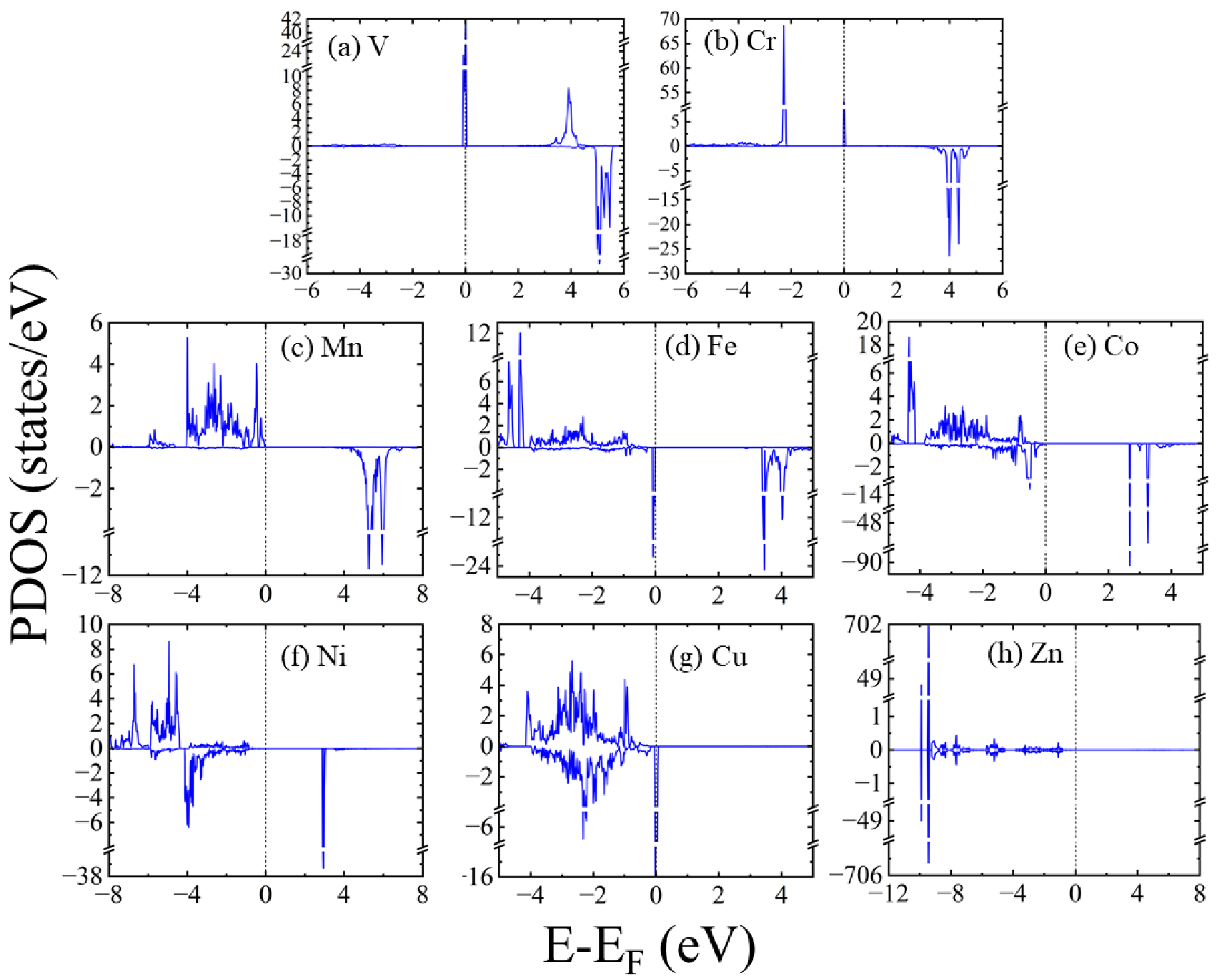}
\caption{Projected DOS on the $3d$ orbitals of TM atom in Pb$_{9}$TM(PO$_4$)$_6$O with the different TM atom in the $4f$ site.
 } \label{fig5}
\end{center}
\end{figure}

The continuous adjustability of electronic structure is a unique feature of TM-doped lead apatite. When the dopant vary from Mn, Fe, Co, Ni, Cu, to Zn in the Pb$_{9}$TM(PO$_4$)$_6$O compound, the local moment around the doped TM atoms decrease continuously from 5.0 to 0 by a step of 1.0 $\mu_B$, despite of the TM atom replacing Pb(1) or Pb(2) atom.
The magnetic moment are listed in Table. \ref{moment}.
To explain the variation of electronic structure including the magnetic moment, we display the projected DOS on $3d$ orbitals of the doped TM atom in the case of replacing Pb(1), shown in Fig. \ref{fig4}.
From Mn to Cu in Fig. \ref{fig4} (a) $\sim$ (e), the electronic states of $3d$ electrons are strongly spin-polarized and their states in the spin-up channel are all occupied. For Mn atom, the spin-down is empty completely, leading to the large moment of 5.0 $\mu_B$. With the element number increasing, more and more spin-down states are filled, which results in the moment decreasing step by step.
As for the doped Zn atom, the spin-up and spin-down $3d$ electronic states are degenerate, corresponding to the nonmagnetic phase.
The above analysis is suitable to the projected DOS in the case of replacing Pb(2) in Fig. \ref{fig5}.

\begin{table}
	\caption{For the Pb$_{9}$TM(PO$_4$)$_6$O with substituting Pb(1) or Pb(2), the magnetic moments of the doped TM atoms are listed. The unit of moment is $\mu_B$.}
	\label{moment}
	\renewcommand\tabcolsep{7.5pt} 
\begin{tabular*}{8.5cm}{ccccccccc}
		\hline
		site&V &Cr & Mn & Fe & Co & Ni & Cu & Zn  \\
		\hline
		$6h$ &3.0 &4.0  &5.0  & 4.0 &3.0  & 2.0 & 1.0 & 0.0 \\
		$4f$ &3.0 &4.0  &5.0  & 4.0 &3.0  & 2.0 & 1.0 & 0.0 \\
		\hline
\end{tabular*}
\end{table}

\subsection{Diluted magnetic ions and weak exchange couplings}

Because the 10\% doping of TM atoms in the Pb$_{10}$(PO$_4$)$_6$O, the TM atoms are separated with a long distance. Considering the large local moment and the moderate size of band gap, we think the TM-doped Pb$_{10}$(PO$_4$)$_6$O is a category of good diluted magnetic semiconductor. Diluted magnetic semiconductors have the properties of both semiconductors and magnetic materials, making it possible to simultaneously utilize electronic charges and spin in semiconductors to fabricate the new electronic devices.
In addition, the diluted magnetic ions may have a close connection with the electrical transport properties, and the TM-doped Pb$_{10}$(PO$_4$)$_6$O can act as a platform for studying the Kondo effect. Even the resistance jump at about 380 K in the Cu-doped lead apatite can be attributed to the resistance minimum due to the Kondo effect at a relatively high temperature.

For Cu-doped lead apatite, the 1 $\times$ 1 $\times$ 2 and 2 $\times$ 1 $\times$ 1 magnetic supercells are built to study the exchange coupling between two Cu moments. The exchange coupling of two Cu atoms in the 1 $\times$ 1 $\times$ 2 supercell or 2 $\times$ 1 $\times$ 1 supercell represents the magnetic coupling along $c$ or $a$ axis direction ($a$ is equal to $b$ axis). With the dopant Cu at $6h$ site, the weak exchange coupling along $c$ direction is less than 0.08 meV, and in the $ab$ plane, the coupling less than 0.005 meV. Therefore, although there is large local moment around the dopant, the coupling between them is too weak to form the stable magnetic order in the Cu-doped Pb$_{10}$(PO$_4$)$_6$O, which make the compound to take on some paramagnetic properties. Other TM-doped lead apatite compounds are similar to the Cu-doped Pb$_{10}$(PO$_4$)$_6$O.

\section{Conclusion}
 By the first-principles calculations, we study the crystal, electronic, magnetic properties of TM-doped lead-apatite. For different TM atoms in Pb$_{9}$TM(PO$_4$)$_6$O, the preferred Wyckoff site, $6h$ or $4f$, is identified. The electronic state near the Fermi energy is dominated by the TM $3d$ and O $2p$ orbitals, and the spin polarization of TM $3d$ electrons result in a local moment. From Mn to Zn, their moments vary continuously with the element number increasing. These moments are separated in the compound with a diluted state and there are very weak interactions between each other. Most of Pb$_{9}$TM(PO$_4$)$_6$O compounds are determined to have a semiconducting ground state with the energy gap larger than 2.0 eV except for 0.9 eV of Cu-doped system. More importantly, the features of flat band and sharp DOS peak near the Fermi are found in the V, Fe, and Cr-doped lead apatite compounds, and similar physical properties like the Cu-doped system can be expected to be measured in them.

\begin{acknowledgments}

This work was supported by the National Natural Science Foundation of China under Grants Nos. 11974217, 12274255, 11974207.
\end{acknowledgments}

\bibliography{Ref}

\end{document}